\documentclass[aps, prx, reprint, superscriptaddress, showkeys]{revtex4-2}
\usepackage{amsmath} 
\usepackage[version=3]{mhchem} 
\ce{}
\usepackage[utf8]{inputenc}
\usepackage[T1]{fontenc}
\usepackage{soul}
\usepackage{bm}
\usepackage{threeparttable}
\usepackage{booktabs}
\usepackage{graphicx}
\usepackage{multirow}
\usepackage{textcomp}
\usepackage[symbol]{footmisc}
\usepackage[dvipsnames]{xcolor}
\usepackage[normalem]{ulem}
\usepackage{xr}
\usepackage{cancel}

\newcommand{\cm}{cm$^{-1}$}
\newcommand{\qim}{$Q_{im}$}

\usepackage[unicode=true, bookmarks=true, bookmarksnumbered=true,
  bookmarksopen=true, bookmarksopenlevel=2, breaklinks=false,
  pdfborder={0 0 1}, backref=false, colorlinks=true, hidelinks
]{hyperref}

\begin{document}

\title[]{Fidelity of Machine Learned Potentials:\\ Quantitative
  Assessment for Protonated Oxalate}

\author{Chen Qu} \email{szquchen@gmail.com} \affiliation{Independent
  Researcher, Toronto, Ontario M9B0E3, Canada} \author{Paul
  L. Houston} \affiliation{Department of Chemistry and Chemical
  Biology, Cornell University, Ithaca, New York 14853, USA} \author{Qi
  Yu} \affiliation{Department of Chemistry, Fudan University,
  Shanghai, 200438, P. R. China } \author{Apurba Nandi}
\affiliation{Department of Chemistry and Cherry L. Emerson Center for
  Scientific Computation, Emory University, Atlanta, Georgia 30322,
  U.S.A.}  \author{Joel M. Bowman} \email{jmbowma@emory.edu}
\affiliation{Department of Chemistry and Cherry L. Emerson Center for
  Scientific Computation, Emory University, Atlanta, Georgia 30322,
  USA.}  \author{Valerii Andreichev, Silvan K\"aser, and Markus
  Meuwly} \affiliation{Department of Chemistry, University of Basel,
  Switzerland} \email{m.meuwly@unibas.ch}

\begin{abstract}
There has been a veritable explosion of methods and software to
perform machine-learned regression on datasets of electronic energies
and forces to develop high-dimensional machine learned potential
energy surfaces (ML-PESs). A major, but not deeply-studied aspect is
how well different ML-PESs represent the same dataset on which they
are trained, beyond the standard fitting precision metrics. Here, this
is examined in detail using several ``stress tests'', for two widely
applied machine-learned potential approaches. One is based on
permutationally invariant polynomial (PIP) linear least square
regression and the other is the message-passing neural network PhysNet
approach.  These potentials and dipole moment surfaces are used in
VSCF/VCI calculations of vibrational energies and wavefunctions.  The
energies from the two PESs are directly compared as are the IR
spectra. In addition, tunneling splittings for the hydrogen transfer
between two equivalent structures are reported from using three
methods: ring polymer instanton theory, diffusion Monte Carlo
simulations, and the $Q_{im}$ path method. These calculations require
the evaluation of on the order of one billion energies that are widely
dispersed in the 15-dimensional configurational space. The two PESs
yield results for these quantities in excellent agreement with each
other.
\begin{description}
\item[Abbreviations] PES, PIP
\end{description}
\end{abstract}


\maketitle

\section{Introduction}
Reliable, quantitatively accurate Potential Energy Surfaces (PESs) are
key for investigating, characterizing, and understanding chemical and
biological systems in the gas and condensed phases based on molecular
simulations.\cite{MM.rev:2019,MMNN2020,Braams09,unke2021machine}
Advances in Machine Learning (ML) methodologies have led to the
development of Machine Learning Potential Energy Surfaces (ML-PESs)
which are now widely used to simulate such systems and become part of
the standard toolbox for studying physical and chemical
processes. Recent applications include the simulation of a water box
with $\approx 12$M atoms and a solid-state system of Cu with 100M
atoms\cite{lu202186} using the DeepMD
model\cite{e2018dpmd}. Similarly, ML-PESs have been used in atomistic
simulations of biological
systems.\cite{unke2024biomolecular,zaverkin2025performance} One of the
largest such simulations was that of the HIV capsid with 44M atoms
using the Allegro
model.\cite{musaelian2023learning,kozinsky2023scaling} On the other
hand, such impressive system sizes are still small compared to what
can be handled using empirical energy functions (billions of atoms for
several million time
steps).\cite{kadau2006molecular,germann2008trillion,shibuta2017heterogeneity,casalino2022breathing,santos2024breaking,ugarte2025scaling}
Depending on the level of theory used for the reference data, the use
of ML-PESs promises to reach new levels of accuracy and realism
compared with all-atom simulations based on empirical and even
physics-enhanced energy functions.\cite{MM.ff:2025}\\

\noindent
Within the broader physical chemistry community, ML-PESs together with
established molecular dynamics (MD) suites and methods become a new
standard
approach.\cite{lahey2020,gastegger2021mlsolv,inizan2023scalable,MM.pycharmm:2023,zinovjev2023electrostatic,zinovjev2024emle,kalayan2024neural}
Likewise, performance and stability challenges for ML-PESs have been
established that explicitly target biomolecular
systems.\cite{poltavsky:2025,poltavsky:2025b} Such studies are
required to pave the way towards confident and competent use of these
new technologies by the broader science community as had been the case
with empirical energy functions and electronic structure methods.\\

\noindent
The profound and lasting impression ML-PESs make on the field of
computer-based approaches applied to the dynamics and reactivity of
chemical, biological and materials systems in the gas- and condensed
phase is also reflected in the large number of computer codes that
have become available over the past decade. A recent review article
presents a sweeping overview of the status of the field up to 2025 and
lists more than 90 software packages.\cite{Jiangrev2025} Two among
this list are used here to address a major, but not deeply-studied,
aspect of ML-PESs.  Namely, how well different approaches represent
the same dataset on which they are trained going well beyond
statistical precision metrics.\\

\noindent
The present work reports results on several ``stress tests'' beyond
fitting precision using a permutationally invariant polynomial (PIP)
PES, based on standard linear regression and a PhysNet neural
network-based potential. The PIP and related methods using PIPs have
been reviewed in a recent Perspective, where it was noted that more
than 200 ML-PESs have been reported using PIPs.\cite{PIPSPERS25} PhysNet
belongs to the family of message passing neural networks and was one
of the early architectures developed in that class of NN-based
approaches. Importantly, the features in PhysNet are continuously
refined and together with a model for the total energy PhysNet also
provides geometry-dependent partial charges. The implementation and
broader scope of PhysNet have also been described in the
literature.\cite{MM.pycharmm:2023,MM.rev:2023,MM.charmm:2024}\\

\begin{figure}
    \centering
    \includegraphics[width=0.6\linewidth]{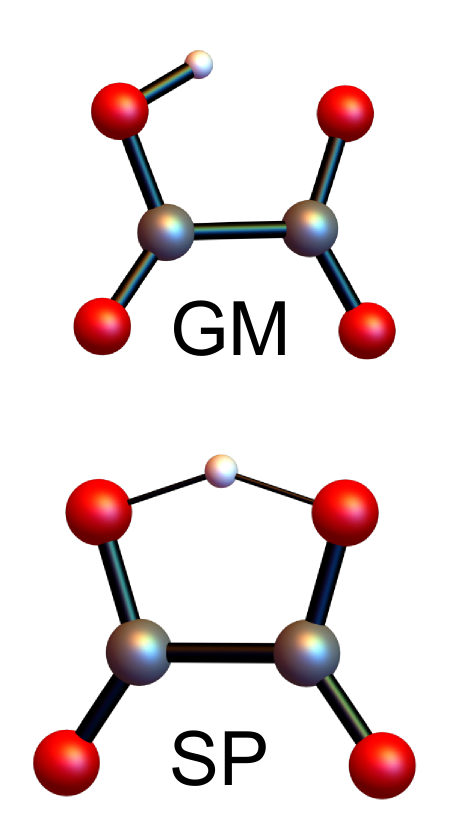}
    \caption{Structures of protonated oxalate anion, OxH, at the
      global minimum (GM) and proton-transfer saddle point (SP).}
    \label{fig:structures}
\end{figure}

\noindent
Specifically, the present work examines the performance of the PIP and
PhysNet ML-PESs for a number of vibrational properties of the
protonated oxalate anion (HO$_2$CCO$_2^-$), hereafter referred to as
OxH, see Figure \ref{fig:structures}, with a major focus on the
infrared (IR) spectrum and on the ground state tunneling
splitting. The splitting was determined using ring polymer instanton
(RPI) theory, unbiased, fixed-node diffusion Monte Carlo (DMC)
simulations, and the 1d $Q_{im}$-path. The remainder of this
manuscript is organized as follows. Brief descriptions of the
theoretical methods are given followed by results, discussion and a
summary and conclusions.\\

\section{Methods}
In the following the data sets and construction of the ML-PESs is
described. Methods for the spectroscopic studies and for computing
tunneling splittings are provided in the supplementary material.\\

\subsection{Reference Data for the PESs}
For the present work two reference data sets were used. Set2025-LL (LL
for ``low level''; MP2/aug-cc-pVTZ) comprised 22100 OxH structures for
which energies, forces, and dipole moments were determined at the
MP2/aug-cc-pVTZ level of theory. To improve the performance of the
LL-model, transfer learning to a ``high level'' (HL) data set
determined at the CCSD(T)/aug-cc-pVTZ level of theory was carried
out.\cite{taylor2009transfer,pan2009survey,smith2019approaching,MM.tl:2022,MM.tlma:2022}
This high-level data set, Set2025-HL, comprised a total of 2688
structures selected semi-randomly from Set2025-LL: 1067 structures
were sampled using MD simulations at 1000~K, and 960 and 661
structures were obtained from normal mode sampling around the minimum
and transition state, respectively, for which energies, forces and
dipole moments were determined at the CCSD(T)/aug-cc-pVTZ level of
theory using MOLPRO.\cite{MOLPRO} A detailed description of the
structural sampling and dataset generation procedures can be found in
previous work.\cite{kaeser2025kernn,MM.oxa:2025} \\

\noindent
For the present work, Set2025-HL was expanded by sampling additional
configurations with a particular focus on improving the tunneling
splitting calculations. For this, 200 points were extracted along the
minimum energy path (MEP) which includes 100 points sampled in each
direction from the transition state (TS) towards the global
minimum. Secondly, 200 points were generated along the instanton path
(IP) generated using PES2025. For all new structures, energy, forces,
and dipole moments were computed at the CCSD(T)/aVTZ level of theory
using MOLPRO.\cite{MOLPRO} This yields Set2026-HL which is the
combination of Set2025-HL and the 400 structures generated along the
MEP and IP.\\

\subsection{PIPs and $\Delta$-Machine Learning}
In general, molecular potentials are invariant with respect to
permutations of like atoms. Permutationally invariant polynomials
(PIPs) provide a direct representation of the potential that respects
this symmetry. In this representation, the potential is given by the
equation
\begin{equation}
V(\bm{\tau})= \sum_{i=1}^{n_p} c_i p_i(\bm{\tau}),
\label{eq1}
\end{equation}
where $c_i$ are linear coefficients, $p_i$ are PIPs, $n_p$ is the
total number of polynomials for a given maximum polynomial order and
$\boldsymbol{\tau}$ are the variables of the potential and the PIPs.
These are Morse variables, denoted $y_{ij}$ and given by $y_{ij}$ =
e$^{-r_{ij}/\alpha}$, where $r_{ij}$ are the $N(N-1)/2$ internuclear
distances among $N$ atoms.\\

\noindent
The general theory underlying this representation was given by Braams
and Bowman\cite{Braams2009}. Two approaches were outlined to obtain
PIPs.  One is given in terms of so-called primary and secondary
invariants and the second is based on ``monomial symmetrization". The
second approach can be illustrated for the simple case of a triatomic.
In that case, the potential is given by
\begin{equation}
V=\sum_{m=0}^M
D_m\mathcal{S}[y_{12}^ay_{13}^by_{23}^c],
\label{eq:symmon}
\end{equation}
where ``$\mathcal{S}$'' is the operator that symmetrizes
monomials. This yields the symmetrized basis function for the
representation of $V$, which for convenience we write in compact
notation of Equation \ref{eq1}.  An efficient algorithm, denoted
``MSA", has been coded to obtain higher order polynomials from lower
order ones.\cite{Xie10, msachen}\\

\noindent
The coefficients, $c_i$, are obtained using linear-least squares
procedure, i.e., they are optimized to minimize the L2 loss, i.e., sum
of the squared differences between the PIP representation of the
potential and the electronic energies. If gradients are also available
these are included in the L2 loss, where the analytical gradient of
$V$ is used.\\

\noindent
In general, the number of coefficients is much less than the dataset
size and so the solution for this overdetermined case makes use of
well-known linear algebra.  We use {dgelss} from the MKL library.  The
routine uses singular value decomposition to obtain the linear
coefficients. Details of the software to generate the PIPs as well as
further processing of them, e.g., fast reverse differentiation using
Mathematica scripts,are given in references
\citenum{PESPIP,msachen}.\\

\noindent
{\it $\Delta$-Machine Learning PIPs:} A $\Delta$-machine learning (ML)
approach to efficiently elevate a low-level PIP ML-PES, e.g., from
extensive DFT or MP2 calculations to the CCSD(T)
level.\cite{DeltaPaper2015,deltaML2021,Nandi2023JACS,JCTCPerspDelta}. In
brief the $\Delta$-ML approach is given by the equation
\begin{equation}
\label{eq:1}
    V_{\rm LL{\rightarrow}CC}=V_{\rm LL}+\Delta{V_{\rm CC-LL}},
\end{equation}
where $V_{\rm LL{\rightarrow}CC}$ is the corrected PES, $V_{\rm LL}$
is a PES fit to low-level reference electronic structure data, and
$\Delta{V_{\rm CC-LL}}$ is the correction PES based on high-level
data, e.g. coupled cluster energies. Since the difference between LL
and HL energies, $\Delta{V_{\rm CC-LL}}$, is not as strongly varying
as $V_{\rm LL}$ with respect to the nuclear configurations and
therefore just a small number of high-level electronic energies are
adequate to fit the correction PES. It is also possible to simply
refit the original low-level dataset by incorporating (I) the
difference energies directly into a new dataset to yield
Set2025-I.\cite{li2022delta,FormAmm2024} The two approaches, as
expected, lead to nearly identical final results.\cite{FormAmm2024}
The latter approach is used here.\\

\subsection{PIP fits to dipole moment data}
Separate fits to dipole moment data are done, using a physically
motivated expression for this vector quantity, namely,
\begin{equation}
    \bm{\mu} = \bm{X} \bm{q},
\end{equation}
where $\bm{X}$ is the $3\times N$ matrix of Cartesian coordinates and
$\bm{q}$ is a $N \times 1$ column vector representing the ``effective
charges'' on the N atoms, which are scalar quantities that can be
expanded using a basis of PIPs to reflect relevant permutational
invariance and with the linear coefficients determined by a linear
least-squares fit. More details can be found in
refs. \citenum{Huang2005,Braams09}.

\subsection{PhysNet and Transfer Learning}
PhysNet is a message-passing neural network that learns to represent
an atom in its chemical environments directly from nuclear charges and
Cartesian coordinates.\cite{PhysNet} The atomic feature vectors are
used to predict atomic energy contributions and partial charges for
any molecular configuration. Forces are obtained through reverse-mode
automatic differentiation\cite{griewank:2008}, while the partial
charges are used to predict the electric dipole moment.\\

\noindent
Following previous work\cite{MM.oxa:2025}, PhysNet's learnable
parameters are adjusted to minimize the loss function
\begin{equation}
  \label{eq:loss}
\begin{aligned}
\mathcal{L} &= w_E \left| E - E^{\text{ref}} \right| + \frac{w_F}{3N}
\sum_{i=1}^{N} \sum_{\alpha=1}^{3} \left| -\frac{\partial E}{\partial
  r_{i,\alpha}} - F^{\text{ref}}_{i,\alpha} \right| \\ &+ w_Q \left|
\sum_{i=1}^{N} q_i - Q^{\text{ref}} \right| + \frac{w_p}{3}
\sum_{\alpha=1}^{3} \left| \sum_{i=1}^{N} q_i r_{i,\alpha} -
p^{\text{ref}}_{\alpha} \right| + \mathcal{L}_{\text{nh}}
\end{aligned}
\end{equation}\\
using the Adam optimizer.\cite{kingma:2014,reddi:2019} The
hyperparameters\cite{MM.physnet:2019,MM.pycharmm:2023} $w_i$ $i \in \{
E, F, Q, p \}$ differentially weigh the contributions to the loss
function and were $w_E = 1$ [1/energy], $w_F \sim 52.92$
[length/energy], $w_Q \sim 14.39$ [1/charge] and $w_p \sim 27.21$
[1/charge/length], respectively, and the term
$\mathcal{L}_{\text{nh}}$ is a ``nonhierarchical penalty'' that
regularizes the loss function.\cite{MM.physnet:2019} For training, a
random 80/10/10~\% split of the data as training/validation/test sets was
used.\\

\noindent
The present work employs two PhysNet-based PESs: PES2025 (available
from previous work\cite{MM.oxa:2025}) and PES2026. For PES2025 first a
low-level (LL) model was trained on Set2025-LL. To improve the quality
of the LL-model for molecular spectroscopy and tunneling splitting
calculations, transfer learning (TL) was used to refine and elevate
the PES from the MP2/aug-cc-pVTZ to the gold-standard
CCSD(T)/aug-cc-pVTZ level of theory using
Set2025-HL.\cite{taylor2009transfer,pan2009survey,smith2019approaching,MM.tl:2022,MM.tlma:2022}
For TL, the learning rate was reduced to $10^{-4}$ and using the
hyperparameters for TL were $w_E = w_F = w_Q = w_p = 1$. This
transfer-learned model is referred to as PES2025 and additional
details are provided in the original publication.\cite{MM.oxa:2025}
For the present work, Set2025-HL was extended in a targeted fashion to
further improve the quality of the tunneling splitting
calculations. Starting from the LL-model based on Set2025-LL (see
above), PES2026 was trained based on Set2026-HL using transfer
learning.\\

\noindent
In addition to the energies, PhysNet also learns an underlying
atom-centered charge model with geometry-dependent charges $q_i$ that
recovers the molecular dipole moment for each configuration. Hence,
the final PhysNet model provides the total molecular energy as the sum
of atomic energies and the total dipole moment $\vec{\mu}(\mathbf{R})
= \sum_{i=1}^{N}q_i(\mathbf{R})\vec{r}_i$ for each geometry, which
yield the full-dimensional potential energy and dipole moment
surfaces. Machine-learned fluctuating charge models, which capture the
nuances of electronic redistribution during nuclear motion, have
recently also been derived from kernel-based approaches and from an
equivariant neural network.\cite{MM.kmdcm.2024,MM.DCMnet.2026} \\

\noindent
The fluctuating charges in PhysNet are directly trained on the
molecular dipole moment $\vec{\mu}$ and are smooth functions of the
geometry. Because these charges are forced to be a smooth function of
the coordinates to minimize the dipole loss, the resulting dipole
moment surface is inherently continuous and differentiable. This is
vital for obtaining high-quality relative intensities in IR
signatures, as the model accurately captures the dipole derivatives
$\frac{\partial \mu}{\partial{Q}}$ across the potential energy
surface. This methodology aligns with previous work on formaldehyde
\cite{mm.vibratingh2co:2020}, which demonstrated that allowing
electrostatics to fluctuate naturally with geometry provides a robust
foundation for reproducing complex spectroscopic features. \\

\section{Results}
\subsection{Fits and Validation of the Machine Learned PESs}
The present work employs three different PESs. The PhysNet-based
PES2025 was previously trained on Set2025-LL and Set2025-HL and used
for spectroscopic applications (VPT2 calculations and MD
simulations). To extend its validity to tunneling splitting
calculations, it was supplemented by additional reference calculations
at the same level of theory along the IP and MEP to give Set2026-HL,
see Methods. Compared with PES2025 (${\rm MAE}(E) = 3.1$ cm$^{-1}$;
${\rm RMSE}(E) = 16.4$ cm$^{-1}$ on Set2025-LL; ${\rm MAE}(E) = 1.7$
cm$^{-1}$; ${\rm RMSE}(E) = 4.2$ cm$^{-1}$ on Set2025-HL), PES2026 on
Set2026-HL for an independent test set features ${\rm RMSE}(E) = 7.8$
cm$^{-1}$, 26.3 cm$^{-1}$/a$_0$ for gradients (compared with 12.0
cm$^{-1}$/a$_0$ for Set2025-HL) and $2.1\times 10^{-3}$ a.u
($5.3\times 10^{-3}$ debye) for dipoles, see Table
\ref{tab:pes_errors}.\\

\begin{table*}[htbp]
\centering
\caption{Comparison of statistical performance for PES2025, PES2026,
  and PIP PES across various datasets. Results on MAE and RMSE for
  energies, forces, and molecular dipole moment are reported as
  indicated together with the data sets the models were trained on.}
\label{tab:pes_errors}
\begin{tabular}{l l l | c c c c}
\hline\hline
Model & Technique & Dataset & MAE($E$) & RMSE($E$) & RMSE($F$) & RMSE($D$) \\
 & & & (cm$^{-1}$) & (cm$^{-1}$) & (cm$^{-1}$/$a_0$) & a.u. (debye) \\
\hline
PES2025 & Base & Set2025-LL & 3.1 & 16.4 & 84.9 & $2.1\times 10^{-3}$ ($5.4\times 10^{-3}$) \\
        & TL & Set2025-HL & 1.7 & 4.1  & 12.0 & $1.0\times 10^{-3}$ ($2.6\times 10^{-3}$) \\
\hline
PES2026 & TL & Set2026-HL & 3.7  & 7.8  & 26.3 & $2.1\times 10^{-3}$ ($5.3\times 10^{-3}$) \\
\hline
PIP PES & $\Delta-$Learn & Set2025-I  & 7.6  & 23.1 & 47.8 & $1.0\times 10^{-3}$ ($2.6\times 10^{-3}$) \\
        & $\Delta-$Learn & Set2025-HL & 12.2  & 31.1 & 65.1 & $1.1\times 10^{-3}$ ($2.7\times 10^{-3}$) \\
\hline\hline
\end{tabular}
\end{table*}

\noindent
Complementary to these PhysNet-based PESs, a PIP-represented
PES\cite{Braams09,PIPSPERS25} was generated using the PES2025-LL data
set. The PIPs used full ``421 symmetry'', indicating that all 4
oxygens permute with one another and the two carbons permute with one
another. The polynomial basis has 9923 polynomials/coefficients.  As
described in the methods section, this fit was then elevated to the
CCSD(T) level using the $\Delta$-ML method and the Set2025-HL data
set. The RMS errors of the final PES on Set2025-I are 23.1 \cm ~for
energies and 47.8 cm$^{-1}$/a$_0$ for gradients, and on Set2025-HL
they are 31.1 cm$^{-1}$ for energies and 65.1 cm$^{-1}$/a$_0$ for
gradients. The dipole moment surface (DMS) using PIPs featured a
training error of $1.0\times 10^{-3}$ a.u. ($2.6\times 10^{-3}$
debye).\\

\begin{figure*}[ht!]
    \centering
    \includegraphics[width=1.0\linewidth]{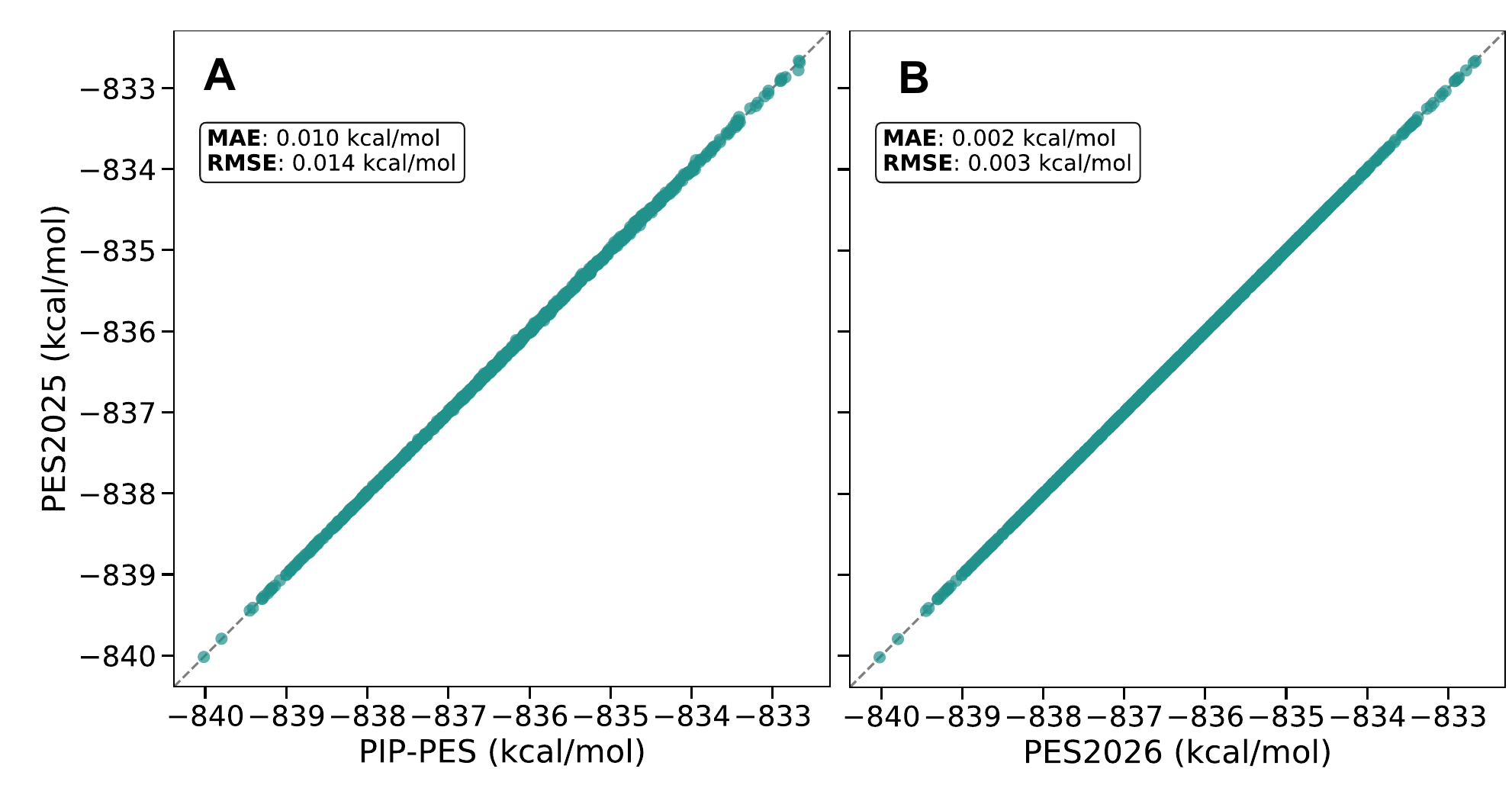}
    \caption{Panel A: Comparison between PES2025 (based on PhysNet)
      and the PIP-based PES. Panel B: Comparison between PES2025 and
      PES2026; for details see text.}
    \label{fig:pes}
\end{figure*}

\noindent
The representations PES2025 and PIP-PES are compared in Figure
\ref{fig:pes}A. For this, 1000 configurations were randomly sampled
from a molecular dynamics (MD) trajectory generated using PES2025 at a
temperature of 300 K. This data set covers a comparatively narrow
energy range of $\sim 10$ kcal/mol. The two PESs feature a MAE$(E)$
and RMSE$(E)$ of 0.010 kcal/mol and 0.014 kcal/mol. Similarly, PES2025
and PES2026 are juxtaposed in Figure \ref{fig:pes}B for which MAE$(E)$
and RMSE$(E)$ are even closer: 0.002 kcal/mol and 0.003 kcal/mol,
respectively. These comparisons demonstrate that all three PESs are
consistent with one another and of very high quality. On the other
hand, the subtle differences should be kept in mind when comparing
observables obtained from using them in the subsequent simulations.\\

\noindent
PES2026 features a barrier height for
H-transfer of 3.349 kcal/mol compared with 3.351 kcal/mol for PES2025
which matches the CCSD(T) reference value. Normal mode frequencies for
the global minimum and the TS were also determined using PES2026 and
their MAE$(\omega)$ matches those from PES2025 to within better than
0.1 cm$^{-1}$. Compared with the frequencies from CCSD(T) calculations
for the minimum energy and TS structure the MAE$(\omega)$ are 0.41
cm$^{-1}$ and 0.55 cm$^{-1}$. All these comparisons underline the high
quality of both, PES2025 and PES2026.\\

\noindent
For the PIP-PES the barrier height for H-transfer from the PIP-PES is
1171.7 cm$^{-1}$, in excellent agreement with the CCSD(T) value of
1172.2 cm$^{-1}$ (3.351 kcal/mol). The harmonic frequencies of these
two stationary points are listed in Table \ref{tab:ccsd_vs_pes}. The
MAE between the reference values and those from the PIP-PES are 0.54
cm$^{-1}$ and 0.92 cm$^{-1}$, respectively. This compares with 0.34
cm$^{-1}$ (CCSD(T)/aug-cc-pVTZ) and 0.09 (MP2/aug-cc-pVTZ) for the
global minimum of PES2025.\cite{MM.oxa:2025}\\

\begin{table*}
    \centering
    \begin{tabular}{cccc|cccc}
        \hline
        \textbf{Min Mode} & \textbf{CCSD(T)} & \textbf{PES} & \textbf{Diff} & \textbf{TS Mode} & \textbf{CCSD(T)} & \textbf{PES} & \textbf{Diff} \\
        \hline
         1 &  104.15 &  104.44 & 0.29 &  1 & 1143.01$i$ & 1141.21$i$ & 1.80 \\
         2 &  300.53 &  300.48 & 0.05 &  2 &  142.08 &  142.68 & 0.60 \\
         3 &  434.85 &  435.22 & 0.37 &  3 &  333.55 &  334.10 & 0.55 \\
         4 &  488.55 &  488.20 & 0.35 &  4 &  492.74 &  493.82 & 1.08 \\
         5 &  565.63 &  565.45 & 0.18 &  5 &  602.51 &  602.91 & 0.40 \\
         6 &  700.53 &  700.64 & 0.11 &  6 &  715.09 &  714.51 & 0.58 \\
         7 &  829.31 &  828.73 & 0.58 &  7 &  750.74 &  749.58 & 1.16 \\
         8 &  849.26 &  849.41 & 0.15 &  8 &  852.74 &  851.81 & 0.93 \\
         9 &  968.64 &  967.52 & 1.12 &  9 &  865.80 &  865.14 & 0.66 \\
        10 & 1140.19 & 1141.01 & 0.82 & 10 & 1296.66 & 1298.50 & 1.84 \\
        11 & 1335.09 & 1335.15 & 0.06 & 11 & 1299.88 & 1299.93 & 0.05 \\
        12 & 1455.66 & 1455.68 & 0.02 & 12 & 1322.91 & 1323.09 & 0.18 \\
        13 & 1726.89 & 1727.56 & 0.67 & 13 & 1748.71 & 1748.66 & 0.05 \\
        14 & 1816.52 & 1816.89 & 0.37 & 14 & 1790.83 & 1791.51 & 0.68 \\
        15 & 3203.91 & 3205.81 & 1.90 & 15 & 2101.26 & 2099.06 & 2.20 \\
        \hline
        \textbf{MAE} & & & 0.47 & & & & 0.85 \\
        \hline
    \end{tabular}
    \caption{Comparison of harmonic frequencies (cm$^{-1}$) from
      CCSD(T) and the PIP-PES. For PES2025 the frequencies have been
      reported previously with MAEs of 0.34~cm$^{-1}$ for the
        minimum and 0.57~cm$^{-1}$ for the TS.\cite{MM.oxa:2025}
      Given the very small overall MAE$(\omega) = 0.1$ cm$^{-1}$
      between PES2025 and PES2026 the results for PES2026 are not
      explicitly reported.}
    \label{tab:ccsd_vs_pes}
\end{table*}

\subsection{Infrared Spectroscopy}
\begin{figure}[ht!]
    \centering
    \includegraphics[width=0.9\linewidth]{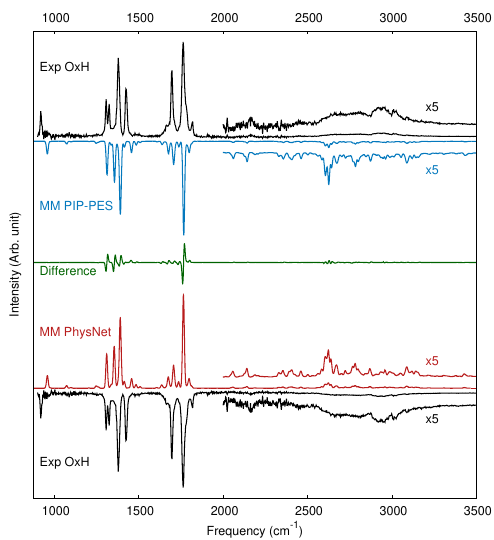}
    \caption{IR spectrum for OxH calculated at the respective global
      minimum using MULTIMODE (MM) and a 4MR with two different
      potential and dipole-moment surfaces. The spectra using the
      PIP-PES and PES2025 are reported in blue and red, respectively,
      whereas the difference between the two computed spectra is shown
      in green. The experimentally measured spectrum is displayed in
      black. Results with a 3MR are reported in Figure
      \ref{sifig:3mr4mr}.}
    \label{fig:ir-MM}
\end{figure}

The computed IR spectra from PES2025 and the PIP-PES, together with
their respective dipole moment surfaces, together with the most recent
experiment are shown in Figure \ref{fig:ir-MM}.\cite{MM.oxa:2025} It
can be seen that the spectra (blue for the PIP-PES and red for
PES2025) from VSCF/VCI calculations using both PESs successfully
capture the location and breadth of the highly dispersed band in the
expected OH-stretch region ranging from $\sim 2500$ cm$^{-1}$ to $\sim
3200$ cm$^{-1}$. A detailed analysis of this band and its implications
for quantum chaos using the PIP-PES has been recently
discussed.\cite{QMchaos} The focus in the present work is on the
fidelity of the potential and dipole moment surfaces that are
constructed with two different approaches (neural network
v.s. polynomial regression) using the same reference data. A direct
comparison of the VSCF/VCI energies from both ML-PESs is given in
Figure \ref{sifig:vscf}.\\

\noindent
The two computed spectra in Figure \ref{fig:ir-MM} (blue and red
traces), are visually identical, even for the fine structures in the
highly dispersed band between 2000 and 3500 \cm. The quantitative
difference between the two computed spectra is the green trace: peak
positions for MM calculations using the two PESs differ by a few
cm$^{-1}$ and arise primarily for the framework modes (1200 cm$^{-1}$
to 1800 cm$^{-1}$). It is also noted that the relative intensities of
the modes are realistic and in reasonable agreement with the
measurements. It is recalled that PhysNet uses fluctuating charges
from which the fluctuating molecular dipole moment is computed whereas
for the PIP-PES the dedicated DMS PIP representation is used.\\

\noindent
The MULTIMODE spectra are compared with the measured ones, starting at
the red side of the spectrum. The weak band below 1000 cm$^{-1}$ is
not correctly described by the MM calculations whereas the quartet
below 1500 cm$^{-1}$ rather closely follows the experiments. The
spacing for the first doublet is somewhat too wide and the intensity
of the highest-frequency peak is too low. For the doublet (experiment)
between 1500 cm$^{-1}$ and 2000 cm$^{-1}$ the MM calculations reveal
more detail for the low-frequency peak whereas the high-frequency peak
is correctly described. Overall, the agreement between the two
computed spectra is excellent and they closely match the
experiments. For the diffuse band extending from 2500 to 3200 cm$^{-1}$
(see black traces in Figure \ref{fig:ir-MM}) the MM calculations using
the two different PESs (PES2025 and PIP-PES) yield nearly identical
results (green trace in Figure \ref{fig:ir-MM}). The computations
using MM confirm that the intensity of the spectroscopic feature(s)
associated with H-transfer are spread out across several hundred
wavenumbers. On the other hand, the computed lineshape does not find
the plateau centered around 2750 cm$^{-1}$.\\

\subsection{Tunneling Splittings}
Next, tunneling splittings from the methods described above are
presented here. Most determinations are for the ground vibrational
state, but some excited state splittings are also reported using the
$Q_{im}$ method.\\

\noindent
{\it Instanton Calculations:} Tunneling splittings had been previously
determined using PES2025 which was transfer-learned from the MP2 level
of theory to CCSD(T). For this PES the training data set was not
specifically tailored to tunneling splitting calculations as had been
done previously for malonaldehyde or
tropolone.\cite{MM.tlma:2022,MM.tl:2025} In order to further improve
PES2025 for tunneling splittings, reference energies at the
CCSD(T)/aVTZ level of theory were determined along the minimum energy
and instanton paths. Subsequently, the TL-procedure was repeated which
yielded PES2026, see Figure \ref{fig:pes}. The tunneling splitting
calculated using perturbatively corrected (pcRPI) RPI theory using
PES2025 was 35.0 cm$^{-1}$. This compares with a value of 34.9
cm$^{-1}$ when employing PES2026. The difference of 0.1 cm$^{-1}$
indicates that both surfaces provide comparable performance and that
the prediction at the CCSD(T)/aVTZ level of theory is $\sim 35.0$
\cm.\\

\noindent
{\it DMC Calculations:} Diffusion Monte Carlo calculations were
performed with both the PIP-PES and the PES2025 surfaces to determine
the zero point and first excited levels of OxH. The latter calculation
was accomplished by preventing the walkers from crossing the nodal
surface. The calculation with the PIP-PES was performed in terms of
normal modes, while that using the PES2025 surface was performed in
Cartesian coordinates. Uncertainties were calculated from the standard
deviation of five simulations. For the zero-point energy, the PIP-PES
gave 7820.1 $\pm$ 1.6 \cm, and the PES2025 surface gave 7819.0 $\pm$
2.0 \cm.  For the first excited levels, the results were 7853.1 $\pm$
0.9 \cm ~and 7856.9 $\pm$ 1.0 \cm, respectively, resulting in
tunneling splittings of 33.0 $\pm$ 1.8 \cm ~and 37.9 $\pm$ 2.3 \cm.\\

\begin{figure}[htbp!]
    \centering
    \includegraphics[width=.7\linewidth]
    {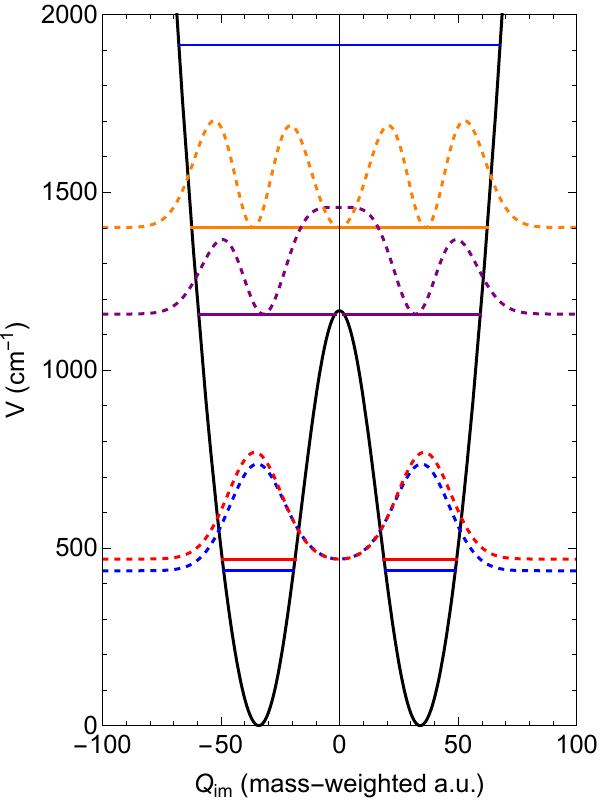}
    \caption{$V(Q_{im}$) potential for OxH (black) with DVR
      energy levels and $|\Psi^2|$ distributions (dashed).}
    \label{fig:Qim}
\end{figure}

\noindent
{\it $Q_{im}$ Calculations in 1d and 2d:} These calculations were all
performed using the PIP-PES. Figure \ref{fig:Qim} reports $V(Q_{im})$
together with the energy levels and probability densities derived from
the wavefunctions. The barrier is 1172 \cm.  Using the normal
coordinate system of the saddle point, the normal coordinate of the
minimum is [34.2, 0.0, 20.2, 0.0, 0.6, 31.0, 10.3, 0.0, -14.2, 0.0,
  -4.5, 3.2, 3.4, 0.1, -8.4] (or, for the other minimum, [-34.2, 0.0,
  20.2, 0.0, -0.6, 31.0, -10.3, 0.0, -14.2, 0.0, 4.5, 3.2, 3.4, -0.1,
  -8.4]), where the 15 numbers represent distances in multiples of the
unit vectors associated with the normal modes of the saddle point,
starting with that of the imaginary frequency and then in order of the
frequencies from lowest to highest (see Table
\ref{tab:ccsd_vs_pes}). \\

\noindent
Energy levels were also calculated for a barrier height of 1250 \cm~
to agree with the CBS limit extrapolated from CCSD(T)-F12a/AVXZ (X=D
and T) calculations, see Table \ref{sitab:qim2050} and Figure
\ref{sifig:qim17722050}. Using the geometries of the \qim ~path for
the 1172 \cm ~barrier, points along the path were calculated at the
CCSD(T)-F12a CBS level and then fit by a polynomial function that was
used in the DVR calculation.  Use of
DVR for this curve gave a tunneling splitting of 28 \cm.\\

\begin{figure}[htbp!]
    \centering
    \includegraphics[width=0.8\linewidth]{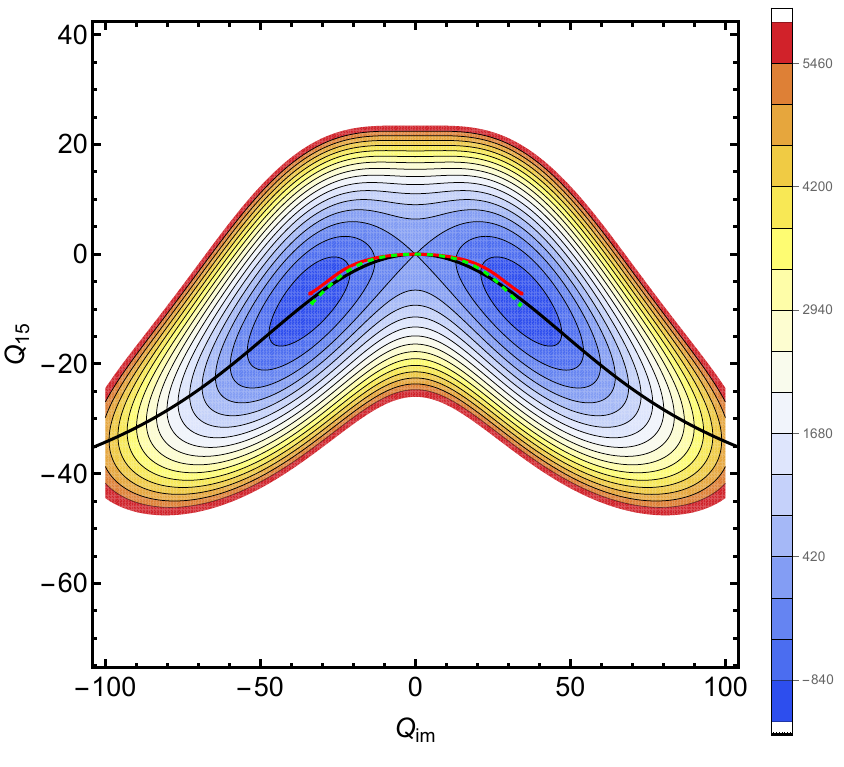}
    \caption{Two dimensional potential energy surface (colored
      contours) for OxH as a function of $Q_{im}$ and $Q_{15}$ with
      all other normal modes optimized at each point using
      PIP-PES. The instanton, minimum energy and $Q_{im}$ path are the
      red, green dashed, and black paths. The instanton and minimum
      energy paths were determined on PES2025 and the $Q_{im}$ path on
      the PIP-PES.}
    \label{fig:QimQ1contour}
\end{figure}

\noindent
A 2d version of the PES was also examined using $Q_{im}$ and $Q_{15}$,
the highest frequency normal mode of the saddle point, as the two
dimensions and then performing an optimization of the remaining $3N-8$
normal coordinates at each value of $Q_{im}$ between --100.0 a$_0$ and
100.0 a$_0$; and at each value of $Q_{15}$ between --72.0 a$_0$ and 40
a$_0$; $Q$ is reported in mass weighted atomic units whereby the mass
is in units of $m_{\rm e}$. The potential surface is shown in colored
contour form in Figure \ref{fig:QimQ1contour}. The grid spacing for
the diagram is 0.5 a$_0$ in each dimension. The black solid line shows
the $Q_{im}$ path on the PIP-PES, the green, dashed line shows the
minimum energy path on PES2025, and the red line shows the instanton
path on PES2025. As can be seen, the three paths are essentially
superimposed in the regions between the transition state and the
global minimum. The associated tunneling splitting is also 33
cm$^{-1}$, nearly identical to that from the 1d \qim~ analysis and in
close agreement with the DMC simulations using the PIP-PES as well as
with the RPI calculations using PES2025. Because the RPI and \qim
~paths pass through the saddle point, corner cutting in OxH is small,
unlike in tropolone, where pronounced corner cutting was
observed.\cite{nandi2023ring} This contributes to the present finding
that $\Delta_{\rm H}$ from the 1d-\qim ~approach, which is simpler and
more approximate than DMC and RPI simulations, compares rather
favorably with these methods and representations of the PESs. A paper
comparing the \qim ~method for a variety of H-transfer molecules has
recently appeared.\cite{qimtests}\\

\begin{table}[htbp!]
    \centering
    \begin{tabular}{l|ccccc}
        \hline
     \textbf{PES} & \textbf{Barrier}& \textbf{Tunneling} &\textbf{Method} & \textbf{Ref.} \\

    & \textbf{\cm} & \textbf{ Splitting (\cm)} &   &  &  \\
        \hline

   PES2025  & 1172 & 35      & RPI & \citenum{MM.oxa:2025} \\
   PES2026 & 1171 & 35      & RPI &  this work \\
   PIP-PES  & 1172 & 36 & DMC & this work \\
   PES2025  & 1172 & 38      & DMC & this work \\
   PIP-PES      & 1172 & 33      & 1d $Q_{im}$  & this work\\
   PIP-PES      & 1172 & 33      & 2d $Q_{im}$  & this work\\
   PIP-PES      & 1250 & 28      & 1d $Q_{im}$  & this work\\
        \hline
    \end{tabular}
    \caption{Comparison of the tunneling splittings $\Delta_{\rm H}$
      for H-transfer in OxH using the different methods described in
      the text.}
    \label{tab:qimtable}
\end{table}

\noindent
{\it Tunneling summary:} A summary of the tunneling splittings for OxH
is provided in Table \ref{tab:qimtable}.  Given the similarities
between the various potential energy surfaces, it is not surprising
that all results using a barrier of 1172 \cm ~give comparable
splittings.  What is more encouraging is that the different methods --
RPI, DMC, \qim ~-- also give consistent results. A comparison of the
instanton and $Q_{im}$ curves on the 2d PIP surface shows that they
are almost identical. This explains why the RPI and $Q_{im}$
calculations agree to within $\sim 2$ cm$^{-1}$, i.e. within $\sim 5$
\%. The corrected barrier height of 1250~\cm ~estimated from the
CCSD(T)/CBS extrapolation and used for a corrected 1d $Q_{im}$ curve
leads to a smaller splitting $\Delta_{\rm H} = 28$ cm$^{-1}$. For a
barrier of 1172 \cm, all of the tunneling splittings are quite
consistently between 33 and 36 \cm. For the 1250~\cm ~barrier, the
splitting is lower. Given the high level of the electronic structure
calculation for that higher barrier, the splitting of 28 \cm ~may
indicate that the actual barrier is in the 28-33 \cm ~range. This is a
testable prediction from highest-level treatments for experimental
verification.\\

\section{Summary and Outlook}
The performance of two machine-learned potentials, $\Delta$-ML PIP and
TL-PhysNet, trained on the same datasets of energies and forces, was
examined for OxH. Dipole moment fits were also done, using two
corresponding approaches. With both representations of the PESs, the
IR spectra and ground-state tunneling splittings were calculated by
using state-of-the-art quantum methods, with nearly identical
results. Each of these calculations - the VSCF/VCI calculations and
the DMC simulations - sample of order $10^9$ configurations. Hence,
this represents a major and quantitative test of the agreement between
the predictions of the two ML-PESs. Notably, each of these
calculations - the VSCF/VCI calculations and the DMC simulations -
sample of order $10^9$ configurations.\\

\noindent
Although not the main point of the paper, it is worth noting that the
width of the highly diffuse and weak OH-stretch band from using both
PESs is consistent with experiment. This also indicates the high
accuracy of the two ML-PESs. However, the fine details are not in
perfect agreement with the measurements. The ground state double-well
tunneling splitting, predicted to be $\sim 30$ cm$^{-1}$, has not been
measured nor explicitly treated in any of the calculations of the IR
spectrum, including the ones shown here. Hence, the influence of the
double-well splitting on the computed IR spectrum is as yet unknown
and remains challenging. Hopefully, this will be uncovered both in
future experiments and calculations. The present PESs are of
sufficient quality to be used in such future studies.\\

\section*{Supplementary Material}
The supplementary material contains a comparison of IR spectra for 3MR
and 4MR VSCF/VCI calculations and details of the CCSD(T)/CBS energies
along the $Q_{im}$-path.

\section*{Data Availability}
The codes and data for the present study are available from
\url{https://github.com/MMunibas/oxalate-2} upon publication.

\section*{Acknowledgment}
The authors gratefully acknowledge financial support from the Swiss
National Science Foundation through grants $200020\_219779$ (MM),
$200021\_215088$ (MM), and the University of Basel (MM). This article
is also based upon work within COST Action COSY CA21101, supported by
COST (European Cooperation in Science and Technology) (to MM).\\

\clearpage

\renewcommand{\thepage}{S\arabic{page}}
\renewcommand{\thetable}{S\arabic{table}}
\renewcommand{\thefigure}{S\arabic{figure}}
\renewcommand{\theequation}{S\arabic{equation}}
\renewcommand{\thesection}{S\arabic{section}}
\setcounter{figure}{0}
\setcounter{section}{0}
\setcounter{table}{0}

\section*{Supporting Material: Fidelity of Machine Learned Potentials: Quantitative Assessment for Protonated Oxalate}

\section{Methods}
\subsection{IR Spectrum Calculations}
VSCF/VCI calculations were performed with the code
MULTIMODE\cite{Multimode1,mmwsp} using the $\Delta$-ML PIP and PES2026
and corresponding full-dimensional dipole moment surfaces. The
interface PyFort, recently reported,\cite{pyfort:2026} was used to
enable MULTIMODE, written in Fortran, to call the PhysNet PES. (The
PIP PES is written in Fortran and so was directly callable in
MULTIMODE.)\\

\noindent
The VSCF ground state and the virtual states form a set of orthonormal
basis functions that can be used for the CI calculation. To restrict
the size of the CI matrix, the excitation space is limited to four
mode excitations (4MR).  The result of these parameters choices are
two H-matrices CI matrices of A' and A'' are 36 674 and 26 408,
respectively. Further details of these calculations can be found in
the Supporting Material of reference \citenum{QMchaos}.\\

\subsection{Diffusion Monte Carlo Calculations}
Unbiased DMC calculations\cite{Anderson1975, kosztin1996introduction}
were performed in two different coordinate systems: the normal
coordinate and the Cartesian coordinate. In both coordinate systems,
the fixed-node DMC\cite{Anderson1976} were employed to calculate the
energy of the first excited state, and regular DMC were used for
ground state.\\

\noindent
In normal coordinates of the saddle point, the nuclear Hamiltonian is
\begin{equation}
    \hat{H} = -\frac{1}{2}\sum_{i=1}^{15}\frac{\partial^2}{\partial
      Q_i^2} + V(\bm{Q}),
\end{equation}
which neglects the vibrational angular momentum terms, for both the
ground and excited state.  This should be an excellent approximation
for this large molecule. The node in normal coordinate is placed at
$Q_{im}=0$, and results with and without the recrossing
correction\cite{Anderson1976} are obtained. This Hamiltonian was used
previously to obtain tunneling splittings for malonaldehyde using
DMC\cite{malon08} and also MCTDH calculations. The splitting was also
calculated using DMC in Cartesian coordinates, where the nodal surface
is not as obvious.\\

\noindent
In each DMC calculation, 30000 walkers were propagated for 55000
steps, with the first 5000 steps for equilibration, and the energies
from the remaining 50000 steps were used to calculate the energy of
the vibrational state. Five DMC calculations were performed for each
state, and the standard deviation of the five simulations are used to
estimate the uncertainty of the DMC calculations.

\subsection{Ring-polymer Instanton Calculations}
Accurate tunneling splittings can be computed using the ring polymer
instanton (RPI) method.\cite{tunnel,InstReview} RPI is a semiclassical
method which relies on locating the optimal tunneling pathway, known
as the instanton, and is defined as an imaginary-time
$\tau\rightarrow\infty$ path connecting two degenerate wells which
minimizes the total action, $S$. The path is constructed by optimizing
a discretized path consisting of $N$ ring-polymer beads and taking the
limit $N\rightarrow\infty$ (typically $N \sim 1000$ is sufficient for
convergence). The potential $U_N$ of a ring polymer is given by
\begin{align}
  \label{eq:rpi_polymer_potential}
    U_N(\bm{x};\beta) = \sum_{i=1}^N V(\mathbf{x}_i) +
    \frac{1}{2(\beta_N\hbar)^2}\sum_{i=1}^N |\mathbf{x}_{i+1} -
    \mathbf{x}_i|^2 \equiv \frac{S({\bm x})}{\beta_N \hbar}
\end{align}
with $\bm{x} = \left( \mathbf{x}_1, \dots, \mathbf{x}_N\right)$ being
the mass-scaled coordinates of the beads, $\beta = \frac{1}{k_b T}$
and $\beta_N = \beta/N$. The first term in
Equation~\ref{eq:rpi_polymer_potential} corresponds to the sum over
all single bead potentials and the second term represents the harmonic
springs with frequency $1/(\beta_N\hbar)$ that connect adjacent
beads. A detailed description of the method is provided,
\textit{e.g.}, in References~\citenum{tunnel} and
\citenum{InstReview}.\\

\noindent
Because of the equivalence $U_N \equiv S$, the action $S$ contains
information \emph{along} the instanton path (IP). Within standard RPI
(sRPI) theory, fluctuations {\it around} the instanton path are
included up to second order and the information is combined into a
contribution $\Phi$. More specifically, fluctuations around the path
are determined from the Hessian matrix at each of the
beads.\cite{richardson2018review,MM.tlma:2022} With this, the
leading-order tunneling splitting in a symmetric double-well system is
\begin{align}
    \Delta_{\rm RPI} = \frac{2\hbar}{\Phi} \sqrt\frac{S}{2\pi\hbar} \,
    \mathrm{e}^{-S/\hbar}.
\label{sieq:rpi_splitting}
\end{align}
One limitation of the sRPI method for determining tunneling splittings
is that fluctuations around the instanton are only included up to
second order.\cite{InstReview} To a good approximation, it is expected
that this captures the dominant tunneling contribution, except for
cases in which anharmonic effects perpendicular to the instanton are
significant or where the barrier is low. For this reason, a
perturbatively corrected RPI (pcRPI) theory was recently
developed\cite{lawrence2023perturbatively}, that accounts for
anharmonicity by including information from the third and fourth order
derivatives of the potential along the instanton. The tunneling
splitting obtained from pcRPI theory is denoted as $\Delta_{\rm PC}$
in the following which, in practice, is obtained from scaling
$\Delta_{\rm RPI}$ with a correction factor $c_{\rm PC}$ according to
$\Delta_{\rm PC} = c_{\rm PC}\cdot \Delta_{\rm RPI}$.\\

\subsection{Q$_{im}$-path Calculations}
In this approach the imaginary-frequency normal mode vector of the saddle
point is used as a rectilinear reaction coordinate. In its simplest
variant, the ``reaction'' path potential $V(Q_{im})$ is the relaxed 1d
minimum energy path along the imaginary-frequency mode of the
H-transfer saddle point.\cite{malon08} It is obtained by varying the
value $Q_{im}$ and optimizing all the remaining $3N-7$ normal
coordinates at each value of $Q_{im}$. Neglecting vibrational angular
momentum,\cite{kamar09} the corresponding Schr\"odinger equation is
\begin{equation}
\label{eq:qim}
\left[-\frac{1}{2}\frac{\partial^2}{\partial
    Q_{im}^2}+V({Q_{im}})\right] \psi_n(Q_{im})=E_n\psi_n(Q_{im})
\end{equation}

\noindent
Note that the minimization is performed starting at the saddle point,
proceeding to minimum, and then beyond the minimum to the repulsive
region, see Figure \ref{fig:Qim}. With this effective 1-dimensional
potential the energies and wavefunction are easily obtained using a
1D-DVR approach.\cite{cmdvr}\\

\section{SM-1: Comparison of VSCF/VCI Energies from the PhysNet and PIP PESs }
\begin{figure}[h]
    \centering
    \includegraphics[width=0.75\linewidth]{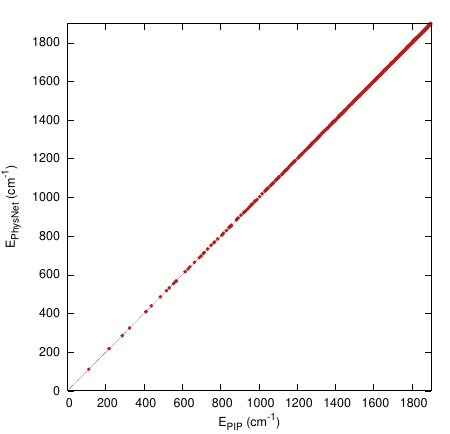}
    \caption{``Correlation" plot of VSCF/VCI energies from PhysNet and
      PIP PESs}
    \label{sifig:vscf}
\end{figure}

\section{SM-2: Comparison of 3MR and 4MR IR spectra using PIP PES and DMS}
\begin{figure*}
    \centering
    \includegraphics[width=0.75\linewidth]{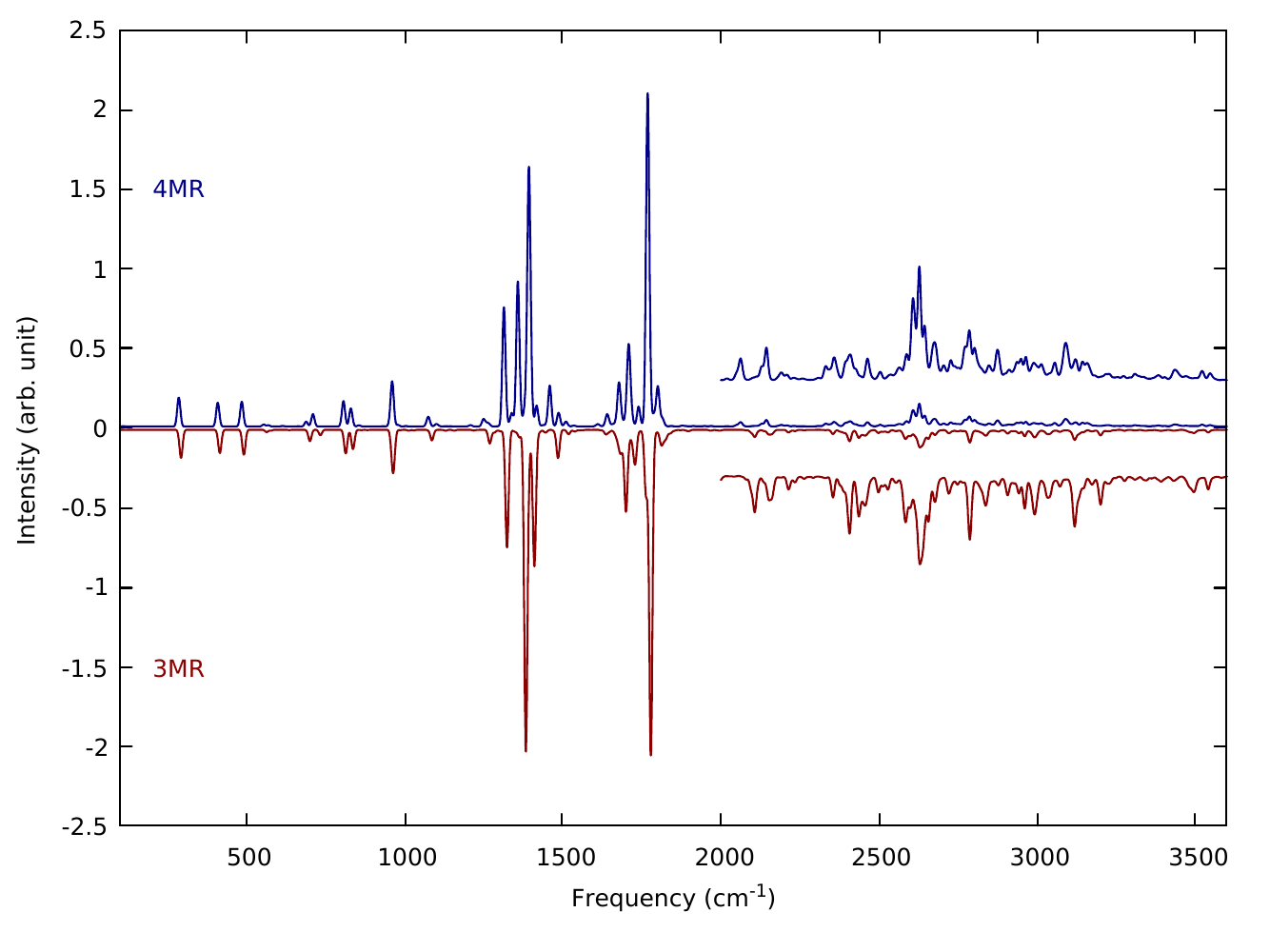}
    \caption{VSCF/VCI IR spectra of protonated oxalate using 4MR and
      3MR of the PIP PES.}
\label{sifig:3mr4mr}
\end{figure*}

Figure \ref{sifig:3mr4mr} compares the IR spectrum obtained with a
3-mode representation of the PIP PES for protonated oxalate with the
4MR one shown in the main text.

\section{SM-3: CBS Results and Calculation for OxH Tunneling Splitting for a Barrier of 1250 \cm}
\begin{figure*}
    \centering
    \includegraphics[width=0.7\linewidth]{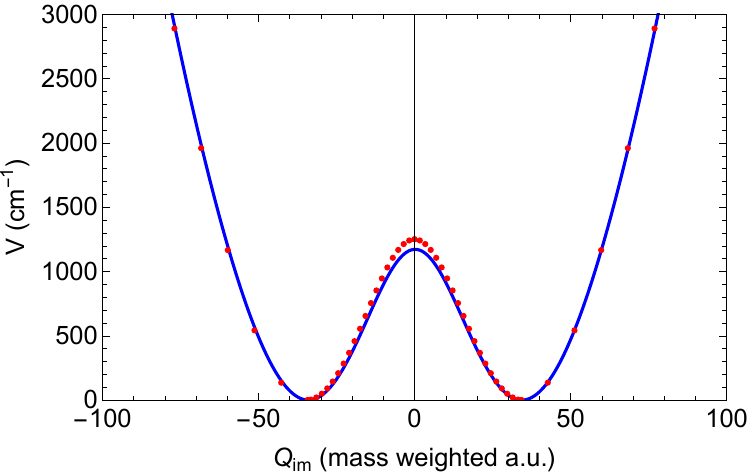}
    \caption{$V(Q_{im})$: (blue line) CCSD(T)/aVTZ \qim ~curve with a
      barrier of 1172 \cm; (red points) CBS energies with a barrier of
      1250 \cm, see Table \ref{sitab:qim2050}.}
\label{sifig:qim17722050}
\end{figure*}

The CBS limit was extrapolated from CCSD(T)-F12a/AVXZ (X=D and T)
calculations. Only positive displacement CBS \qim ~results are needed
for geometries calculated from the CCSD(T)/AVTZ \qim ~curve.  These
are shown in Table \ref{sitab:qim2050}. Figure \ref{sifig:qim17722050}
compares these results (red points) to the CCSD(T)/AVTZ results (blue
line).  The major difference between the two is that the CBS potential
has a barrier of 1250 \cm, whereas the CCSD(T)/AVTZ barrier is 1172
\cm. \\

\begin{table}[h!]
    \centering
    \begin{tabular}{cc|cc}
        \hline
        \qim  & $V$ (\cm) & \qim & $V$ (\cm) \\
        \hline
 0. & 1250.13 & & \\
 1.70018 & 1241.07 & 27.9534 & 90.8098 \\
 3.45034 & 1213.17 & 29.7036 & 49.193 \\
 5.20051 & 1167.59 & 31.4538 & 20.038 \\
 6.9507 & 1106.19  & 33.2041 & 3.61036 \\
 8.70089 & 1031.33 & 34.2042 & 0. \\
 10.4511 & 945.749 & 42.7052 & 135.185 \\
 12.2013 & 852.367 & 51.2561 & 541.933 \\
 13.9515 & 754.143 & 59.8069 & 1165.57 \\
 15.7017 & 653.949 & 68.3575 & 1958.58 \\
 17.452 & 554.487  & 76.9081 & 2889.29 \\
 19.2022 & 458.221 & 85.4587 & 3937.51 \\
 20.9524 & 367.33  & 94.0093 & 5087.66 \\
 22.7027 & 283.708 & 102.56 & 6326.77 \\
 24.4529 & 208.931 & 111.11 & 7644.85 \\
 26.2031 & 144.294 & 119.661 & 9033.98 \\
    \hline
    \end{tabular}
    \caption{$V(Q_{im})$ for the CBS calculations.}
\label{sitab:qim2050}
\end{table}

\noindent
The CCSD(T)/AVTZ and the CBS results were fit to the even terms of a
20-order polynomial in \qim, and those fits were used to calculate the
energy levels using DVR\cite{cmdvr} with a grid spacing of 0.2 and a
maximum energy of 0.025 a.u., resulting in a grid of 1041 points. For
the CBS curve, lowest two eigenvalues gave a splitting of 28.46 \cm,
as to compared to 32.87 \cm~ for the CCSD(T)/AVTZ curve.\\

\newpage
\clearpage

\bibliography{refs,refsjmb}
\end{document}